\pdfoutput=1

\documentclass[10pt, conference, compsocconf]{IEEEtran}

\let\labelindent\relax

\usepackage{enumitem}

\usepackage{color}
\usepackage{soul}
\usepackage{algorithmicx}
\usepackage{graphicx}
\usepackage{amssymb}
\usepackage{amsmath}
\usepackage{xspace}
\usepackage{subfigure}

\usepackage{algorithm}
\usepackage[noend]{algpseudocode}
\usepackage{hyperref}

\let\OLDthebibliography\thebibliography
\renewcommand\thebibliography[1]{
  \OLDthebibliography{#1}
  \setlength{\parskip}{0pt}
  \setlength{\itemsep}{0pt plus 0.3ex}
}

\floatname{algorithm}{Algorithm}
\renewcommand{\algorithmicrequire}{\textbf{Input:}}
\renewcommand{\algorithmicensure}{\textbf{Output:}}

\newcommand{\vtxs}{{\tt vtxs}}
\newcommand{\nbor}{{\tt nbor}}
\newcommand{\nets}{{\tt nets}}

\newcommand{\CP}{{\tt ColPack}\xspace}

\usepackage{algorithm}
\usepackage{algpseudocode}
\usepackage{pbox}

\begin{document}
\everypar{\looseness=-1}

\title{Greed is Good: Optimistic Algorithms for Bipartite-Graph Partial Coloring on Multicore Architectures}

\author{\IEEEauthorblockN{Mustafa Kemal Ta\c{s}}
\IEEEauthorblockA{Computer Science and Engineering,\\
Sabanc{\i} University, Istanbul, Turkey\\
\small\tt mkemaltas@sabanciuniv.edu}
\and
\IEEEauthorblockN{Kamer Kaya}
\IEEEauthorblockA{Computer Science and Engineering,\\
Sabanc{\i} University, Istanbul, Turkey\\
Dept. Biomedical Informatics,\\
The Ohio State University, Columbus, USA\\
\small\tt{kaya@sabanciuniv.edu}}
\and
\IEEEauthorblockN{Erik Saule}
\IEEEauthorblockA{Computer Science,\\
The University of North \\Carolina at Charlotte,\\
Charlotte, NC, USA\\
\small\tt{esaule@uncc.edu}}
}

\maketitle

\begin{abstract}
In parallel computing, a valid graph coloring yields a lock-free processing of the colored tasks, data points, etc., without expensive synchronization mechanisms. However, coloring is not free and the overhead can be significant. In particular, for the bipartite-graph partial coloring~(BGPC)  and distance-2 graph coloring~(D2GC) problems, which have various use-cases within the scientific computing and numerical optimization domains, the coloring overhead can be in the order of minutes with a single thread for many real-life graphs. 

In this work, we propose parallel algorithms for bipartite-graph partial coloring on shared-memory architectures. Compared to the existing shared-memory BGPC algorithms, the proposed ones employ greedier and more optimistic techniques that yield a better parallel coloring performance. In particular, on 16 cores, the proposed algorithms perform more than $4\times$ faster than their counterparts in the \CP library which is, to the best of our knowledge, the only publicly-available coloring library for multicore architectures. In addition to BGPC, the proposed techniques are employed to devise parallel distance-2 graph coloring algorithms and similar performance improvements have been observed. Finally, we propose two costless balancing heuristics for BGPC that can reduce the skewness and imbalance on the cardinality of color sets (almost) for free. The heuristics can also be used for the D2GC problem and in general, they will probably yield a better color-based parallelization performance especially on many-core architectures. 
\end{abstract}


\IEEEpeerreviewmaketitle

\section{Introduction}
A coloring on a graph $G = (V, E)$ explicitly partitions the vertices in $V$ into a number of disjoint subsets such that two vertices $u, v \in V$ that are in the same color set are independent from each other, i.e., $(u, v) \notin E$. Graphs have been frequently used to model data, e.g., matrices and tensors, as well as computations. In these models, two neighbor vertices usually imply a potential race-condition in a parallel execution. On the other hand, given a valid coloring on $V$, each color set, formed by independent vertices, can be simultaneously processed in a lock-free manner and without a synchronization overhead. Moreover, in practice, a {\em good} coloring with a small number of colors will probably yield a better performance compared to a {\em bad} coloring with a large number of colors since the number of barriers (between the color sets), and hence the parallelization overhead, will be less. Unfortunately, {\em the distance-1 graph coloring problem}~(D1GC), i.e., coloring a graph with the minimum number of colors such that all adjacent vertices have different colors, is NP-Complete and hard to approximate~\cite{matula_SL,Zuckerman}.\looseness=-1

The naive, adjacency-based neighborhood is not suitable for numerous applications such as numerical optimization and efficient computation of Hessians and Jacobians. Instead, the problem can be modeled as a {\em bipartite graph partial-coloring}~(BGPC) problem. In BGPC, given a bipartite graph $G = (V_A \cup V_B, E)$, one wants to color the vertices in $V_A$ with minimum colors, such that all vertex pairs that are adjacent to at least one $V_B$ vertex have different colors. A similar problem is {\em distance-2 graph coloring}~(D2GC), where a graph is colored in a way that the color of each vertex is different than the colors of the vertices in its distance-2 neighborhood. For more details on the applications of BGPC and D2PC and parallel algorithms to solve these problems on shared-memory and distributed-memory architectures, we refer the reader to~\cite{ColPack, CM83, Bozdag05-HPCC, Bozdag10-SISC, GMP05, Gebremedhin02paralleldistance-k}.\looseness=-1

From the parallel computing perspective, another desirable property of a good coloring is the {\em balance} on the color set cardinalities~\cite{LuBalanced,Meyer,Hajnal,Gjertsen}; a more balanced coloring can improve the convergence speed and the value of the final objective function for some iterative algorithms. However, a tight balance is not required if shared-memory parallelism is the only concern; if all the color set cardinalities are above a certain threshold, that depends on the number of processors/cores available and the task heterogeneities, the parallel performance will not be disrupted by the remaining imbalance since there will be enough work to feed all the available cores/processors.\looseness=-1

Good colorings are not free and their generation adds an overhead for parallelization. Furthermore, the impact of this overhead increases if the coloring is performed sequentially and the actual job is executed on a large number of cores. This is why parallelization of graph coloring algorithms have been extensively studied for all the problems above, e.g.,~\cite{ColPack,Bozdag10-SISC,Catalyurek12-ParCo,Deveci2016}. The results in the literature show the execution time of a sequential D1GC algorithm is less than a second for many real-life graphs. However, for D2GC and BGPC, the overhead can be in the order of minutes.\looseness=-1

The contribution of this paper are three-fold: {\bf 1)} We propose parallel BGPC algorithms on multicore architectures that employ greedier and more optimistic techniques compared to the existing algorithms. We compared the performance of the proposed algorithms with the one in the \CP library which, to the best of our knowledge, is the only publicly-available coloring library with a parallel BGPC implementation. On the average for eight UFL matrices and with 16 threads/cores, $1.47\times$ speedup is obtained via basic optimizations, another $2.81\times$ speedup is obtained by employing faster and more optimistic techniques without a significant increase on the number of colors. Overall, the proposed algorithm is $4.71\times$ faster than the parallel \CP implementation on 16 threads and uses only $8\%$ more colors. {\bf 2)} We applied the same techniques for the D2GC problem and  observed similar speedups on the five of eight, square, structurally symmetric matrices in our test-bed. {\bf 3)} We integrated two online heuristics to the proposed BGPC algorithms that aim to balance the color set cardinalities during the course of the coloring without a significant computational overhead: The first heuristic tries not to increase the number of colors, whereas the second one aggressively improves the balance by using more colors (only $11\%$ more on average for the eight graphs in our experiments).

The rest of the paper is organized as follows: Section~\ref{sec:back} introduces the notation and background on parallel coloring algorithms. The proposed 
BGPC algorithms are described in detail in Section~\ref{sec:BGPC} and their adaptation for D2GC is presented in Section~\ref{sec:D2GC}. The balancing heuristics are described in Section~\ref{sec:bal}. Section~\ref{sec:exp} presents the experimental results and Section~\ref{sec:related} briefly surveys the related coloring literature. Section~\ref{sec:conc} concludes the paper.\looseness=-1

\section{Background and Notation}\label{sec:back}

Most of the recent coloring algorithms use a speculative, iterative approach which first colors the vertices optimistically in parallel hoping that a valid coloring will be generated, e.g.,~\cite{ColPack,Catalyurek12-ParCo,Deveci2016,Sariyuce:2012}. The validity of the coloring is then verified in a conflict removal step; if a conflict, i.e., {\em a pair of neighbor vertices with the same color}, is detected, one of the vertices is tagged to be colored in the next iteration. Let $G = (V, E)$ be a graph and let $V_{color} \subseteq V$ be the vertices that need to be colored. Let $\nbor$($v$) $\subset V_{color}$ define the neighborhood structure of the vertices to be colored. Throughout the text, non-negative integers will be used for coloring and {\bf -1} is used for an uncolored vertex. A pseudocode of the greedy optimistic graph coloring approach is given in Algorithms~\ref{alg:opt},~\ref{alg:color} and~\ref{alg:conf}.\looseness=-1
 
\renewcommand{\baselinestretch}{0.9}
\begin{algorithm}
\caption{\textsc{GreedyGraphColoring}}
\algorithmicrequire{ $G = (V, E)$, $V_{color} \subseteq V$: vertices to be colored, $\nbor$($.$): the neighborhood function for the vertices in $V_{color}.$}\\
\algorithmicensure{ $c[.]$:  a valid coloring array for $V_{color}$}
\begin{algorithmic}[1]
\State{$W \gets V_{color}$}\label{ln:fillwq}
\State{$c[v] \leftarrow {\mathbf {-1}}$, $\forall v \in V_{color}$}
\While{$W$ is not empty}\label{ln:wqnotempty}
\State{{$c \gets$ \sc ColorWorkQueue($G$, $W$, $c$)}}\label{ln:colorwq}
\State{$W \gets ${\sc RemoveConflicts($G$, $W$, $c$)}}\label{ln:removeconflicts}
\EndWhile
\end{algorithmic}
\label{alg:opt}
\end{algorithm}

\begin{algorithm}
\caption{\textsc{ColorWorkQueue}}
\algorithmicrequire{ $G = (V, E)$, $W$: vertices to color, $\nbor$($.$): the neighborhood function, $c[.]$: an incomplete coloring with no conflicts.}\\
\algorithmicensure{ $c[.]$:  an optimistic coloring.}
\begin{algorithmic}[1]
\For{each $w \in W$ {\bf in parallel}} 
\State{$F \leftarrow \emptyset$} \Comment{{\bf{thread private}} forbidden color set for $w$}
\For{each $u \in$ $\nbor$($w$)}
\If{$c[u] \neq {\mathbf {-1}}$}
\State{$F \leftarrow F \cup \{c[u]\}$}
\EndIf
\EndFor
\State{$col \leftarrow 0$} \Comment{first-fit coloring policy}
\While{$col \in F$}
\State{$col \leftarrow col + 1$}
\EndWhile
\State{$c[w] \leftarrow col$}
\EndFor
\end{algorithmic}
\label{alg:color}
\end{algorithm}

\begin{algorithm}
\caption{\textsc{RemoveConflicts}}
\algorithmicrequire{ $G = (V, E)$: the graph to color, $W$: vertices to color, $\nbor$($.$): the neighborhood function, $c[.]$:  an optimistic coloring.}\\
\algorithmicensure{ $W_{next}$:  the work queue for next iteration, $c[.]$:  a (probably incomplete) coloring with no conflicts.}
\begin{algorithmic}[1]
\State{$W_{next} \gets \emptyset$}  \Comment{a {\bf{shared}} queue for the next iter. }
\For{each $w \in W$ {\bf in parallel}} 
\For{each $u \in$ $\nbor$($w$)}
\If{$c[u] = c[w]$ and $w > u$}
\State{$W_{next} \leftarrow W_{next} \cup \{w\}$ {: {\bf{atomic}}}}
\State{{\bf break}}
\EndIf
\EndFor
\EndFor
\end{algorithmic}
\label{alg:conf}
\end{algorithm}
\renewcommand{\baselinestretch}{1}

As the algorithms show, at each iteration, a set of vertices in $W$ are optimistically colored. A conflict removal phase is the performed to check if they are conflicting with the other vertices in $V_{color}$. When conflicts are detected, the {\em conflicting vertices} are added to the next iteration's vertex queue and the procedure is repeated. This greedy and optimistic approach can be used for almost all the coloring variants and the definitions of $V_{color}$ and $\nbor$($.$) change with respect to the problem. For the BGPC problem on a bipartite graph $G = (V, E)$ where $V = V_A \cup V_B$ has two parts, $V_{color} = V_A$ and for each $u \in V_A$, $\nbor$($u$) is defined as $\{v \in V_A \setminus \{u\} : \exists w \in V_B$ s.t. $(u,w) \in E$ and  $(v, w) \in E\}$. For D2GC, $V_{color} = V$ and $\nbor$($u$) is the set of vertices in $V$ whose shortest-path distances to $u$ are less than or equal to two.\looseness=-1

The BGPC problem can also be considered as a hypergraph coloring problem~\cite{Bozdag10-SISC} where the elements of $V_A$ correspond to the {\em pins} to be colored, and the ones in $V_B$ correspond to the {\em nets} in the hypergraph which define the neighborhood. Based on this analogy, for clarity, while describing our BGPC algorithms we will use the terms {\em vertex} and {\em net} to denote a $V_A$ and $V_B$ vertex, respectively, in the bipartite graph. Similarly, for a vertex $u \in V_A$~($v \in V_B$), $\nets$($u$)~($\vtxs$($v$)) will denote the set of $V_B$~($V_A$) vertices adjacent to $u$~($v$).\looseness=-1
 
\section{Algorithms for bipartite-graph coloring}\label{sec:BGPC}
In BGPC, both of the coloring and conflict removal phases can be performed in two ways: {\it vertex-based} and {\it net-based}. The existing literature on shared-memory bipartite-graph partial coloring algorithms follow the former approach. However, net-based coloring can be more efficient since the neighborhood single-handedly defines the validity of the coloring. Furthermore, depending on the iteration number and the size of the current work queue $W$, i.e., the number of remaining vertices to be colored, this approach can be more efficient..\looseness=-1

The vertex-based BGPC approach, which is employed by the \CP library, traverses the neighborhood starting from the vertices to be colored both for \textsc{ColorWorkQueue} and \textsc{RemoveConflicts} as shown in Algorithms~\ref{alg:bgpc:v:color} and~\ref{alg:bgpc:v:conf}, respectively.\looseness=-1
 
\renewcommand{\baselinestretch}{0.9}
\begin{algorithm}
\caption{\textsc{BGPC-ColorWorkQueue-Vertex}}
\algorithmicrequire{ $G = (V_A \cup V_B, E)$: a bipartite graph, $W$: vertices to color, $c[.]$: an incomplete coloring with no conflicts.}\\
\algorithmicensure{ $c[.]$:  an optimistic coloring.}
\begin{algorithmic}[1]
\For{each $w \in W$ {\bf in parallel}} 
\State{$F \leftarrow \emptyset$ : {{\bf{thread private}} forbidden color set for $w$}}
\For{each $v \in$ $\nets$($w$)}
\For{each $u \in$ $\vtxs$($v$) $\setminus \{w\}$}
\If{$c[u] \neq {\mathbf {-1}}$}
\State{$F \leftarrow F \cup \{c[u]\}$}
\EndIf
\EndFor
\EndFor
\State{$\ldots$} \Comment{first-fit coloring (lines 6-9 in Alg. 2)}
\EndFor
\end{algorithmic}
\label{alg:bgpc:v:color}
\end{algorithm}

\begin{algorithm}
\caption{\textsc{BGPC-RemoveConflicts-Vertex}}
\algorithmicrequire{ $G = (V_A \cup V_B, E)$, $W$: vertices to color, $\nbor$($.$): the neighborhood function, $c[.]$: an optimistic coloring.}\\
\algorithmicensure{ $W_{next}$:  the work queue for next iteration, $c[.]$: a (probably incomplete) coloring with no conflicts.}
\begin{algorithmic}[1]
\State{$W_{next} \gets \emptyset$  : {a {\bf{shared}} queue for the next iter. }}
\For{each $w \in W$ {\bf in parallel}} 
\For{each $v \in$ $\nets$($w$)}
\For{each $u \in$ $\vtxs$($v$) $\setminus \{w\}$}
\State{$\ldots$} \Comment{detect conflicts (lines 4-6 in Alg. 3)}
\EndFor
\EndFor
\EndFor
\end{algorithmic}
\label{alg:bgpc:v:conf}
\end{algorithm}
\renewcommand{\baselinestretch}{1}

For \textsc{BGPC-ColorWorkQueue-Vertex}, the vertex-based approach needs to go over all the vertices in $V_{color}$ in the first iteration. That is each net $v \in V_B$ will be visited $\vtxs$($v$) times and for each visit, all $\vtxs$($v$) edges will be processed; hence, the complexity of the neighborhood traversal in the first iteration is  $\Theta\left(\sum_{v \in V_{B}}{|\vtxs(v)|^2}\right)$. The complexity of the conflict removal phase for the first iteration is also $\mathcal{O}\left(\sum_{v \in V_{B}}{|\vtxs(v)|^2}\right)$. Although there can be early terminations~(line 6 of Alg. 3), this worst-case bound is tight;  if the optimistic coloring is valid, the neighborhood needs to be traversed for each vertex in $V_{A} = V_{color}$ in the first conflict removal phase. Unfortunately, for many BGPC use cases, such as numerical optimization, there can be $V_B$ nets having tens of thousands of adjacent vertices. These nets will be problematic while coloring a bipartite graph especially for the first iteration that dominates the overall execution time according to our experience.\looseness=-1

In net-based coloring, the vertices are colored by observing the neighborhood from the nets' side; in BGPC, a conflict is created when ``two vertices in the same $\vtxs$\xspace set are colored with the same color". Hence, the net-based approach sounds more natural for coloring. 
The coloring and conflict removal phases of the most straightforward and the most optimistic net-based BGPC are given in Algorithms~\ref{alg:bgpc:n:color:ff} and~\ref{alg:bgpc:n:conf}, respectively.\looseness=-1

\renewcommand{\baselinestretch}{0.9}
\begin{algorithm}
\caption{\textsc{BGPC-ColorWorkQueue-Net-v1}}
\algorithmicrequire{ $G = (V_A \cup V_B, E)$: a bipartite graph.}\\
\algorithmicensure{ $c[.]$:  the (most) optimistic coloring array.}
\begin{algorithmic}[1]
\For{each $v \in V_B$ {\bf in parallel}} 
\State{$F \leftarrow \emptyset$ : {{\bf{thread private}} forbidden color set for $v$}}
\State {$col \leftarrow 0$ : {\bf{thread private}}}
\For{each $u \in$ $\vtxs$($v$)}\label{ln:kk:1}
\If{$c[u] = {\mathbf {-1}}$ {\bf or} $c[u] \in F$}
\While{$col \in F$}\label{ln:kk:3}
\State{$col \leftarrow col + 1$}
\EndWhile
\State{$c[u] \leftarrow col$}\label{ln:kk:2}
\EndIf
\State{$F \leftarrow F \cup \{c[u]\}$}
\EndFor
\EndFor
\end{algorithmic}
\label{alg:bgpc:n:color:ff}
\end{algorithm}

\begin{algorithm}
\caption{\textsc{BGPC-RemoveConflicts-Net}}
\algorithmicrequire{ $G = (V_A \cup V_B, E)$: a bipartite graph to color, $c[.]$: an optimistic coloring.}\\
\algorithmicensure{ $c[.]$: an incomplete coloring.}
\begin{algorithmic}[1]
\For{each $v \in V_B$ {\bf in parallel}} 
\State{$F \leftarrow \emptyset$ : {{\bf{thread private}} forbidden color set for $v$}}
\For{each $u \in$ $\vtxs$($v$)}
\If{$c[u] \neq {\bf -1}$}
\If{$c[u] \in F$}
\State{$c[u] \leftarrow \mathbf {-1}$}
\Else
\State{$F \leftarrow F \cup \{c[u]\}$}
\EndIf
\EndIf
\EndFor
\EndFor
\end{algorithmic}
\label{alg:bgpc:n:conf}
\end{algorithm}
\renewcommand{\baselinestretch}{1}

The net-based coloring in Algorithm~\ref{alg:bgpc:n:color:ff} processes the nets in parallel to color the vertices in the adjacency lists. The complexity of each iteration is linear in terms of the size of the graph $(|V_A \cup V_B| + |E|)$. However, while coloring, each thread only checks the local conflicts within the neighborhood of the current net's adjacency; this is the optimism. When a vertex $u$ is visited~(line~\ref{ln:kk:1}), the thread first checks the value of $c[u]$. If $c[u]$ is not set yet~(or set to {\bf -1} in the previous conflict removal phase), or if $c[u]$ has been used for the current net before, $u$ is recolored~(line~\ref{ln:kk:2}). While doing that, Algorithm~\ref{alg:bgpc:n:color:ff} imitates a net-level first-fit coloring~(lines~\ref{ln:kk:3}--\ref{ln:kk:2}) for the visited vertices. This is the most optimistic net-based coloring since the threads ``hope" that they are using a color in the earlier positions of the adjacency list which will not appear later positions. Unfortunately, our preliminary experiments show that this level of optimism is maleficent due to the large number of conflicts it incurs.\looseness=-1
 
Although the net-based approach is not straightforward to employ for the coloring phase, it suits much better for the conflict removal phase; a net-based traversal given in Algorithm~\ref{alg:bgpc:n:conf} is sufficient to detect all the existing conflicts. Moreover, unlike its vertex-based variant, the complexity of an iteration is linear in terms of the graph size. One drawback is that it may remove more colorings than required compared to vertex-based approach. However, we did not observe a significant performance reduction due to this optimism of net-based conflict detection.\looseness=-1
   
To keep the coloring process in the right track by reducing the number of conflicts, we propose a less optimistic version of \textsc{BGPC-ColorWorkQueue-Net-v1} as in Algorithm~\ref{alg:bgpc:n:color}. There are two main modifications: first, to reduce the number of re-colorings within the adjacency list of a single net, the algorithm first performs a pass on the adjacency list and marks the forbidden colors~(the for loop at line~\ref{ln:kk:01}). While doing that, it also stores the vertices that need to be colored in a thread private queue $W_{local}$~(line~\ref{ln:kk:02}). After the first pass, the vertices in $W_{local}$ are visited and colored one-by-one.\looseness=-1

The second modification is applied while coloring these vertices; instead of using a first-fit policy that uses the smallest possible color for a vertex, we employ a reverse first-fit policy~(lines~\ref{ln:rff1}--\ref{ln:rff3}) that uses the largest possible color smaller than $|\vtxs(v)|$ while coloring the vertices in $W_{local}$. This policy never uses a negative color since there are at most $|\vtxs(v)|$ vertices in $W_{local}$ and $|\vtxs(v)|$ colors can still be used for them. Besides, since $|\vtxs(v)|$ is a lower-bound on the number of colors used, we do not expect a large increase on the number of colors used. Moreover, reverse first-fit is expected to produce less number of conflicts compared to the first-fit in Algorithm~\ref{alg:bgpc:n:color:ff}, since it does not use always use the same small colors but prioritize different colors for each net. To understand the benefits of these modifications better, we refer the reader to Table~\ref{tab:prelim} where the number of colored (remaining) vertices after the first iteration is presented for two graphs when Algorithms~\ref{alg:bgpc:n:color:ff} and~\ref{alg:bgpc:n:color} are employed.\looseness=-1
 
\renewcommand{\baselinestretch}{0.9}
\begin{algorithm}
\caption{\textsc{BGPC-ColorWorkQueue-Net}}
\algorithmicrequire{ $G = (V_A \cup V_B, E)$: a bipartite graph, $c[.]$: an incomplete coloring.}\\
\algorithmicensure{ $c[.]$:  an optimistic coloring array.}
\begin{algorithmic}[1]
\For{each $v \in V_B$ {\bf in parallel}} 
\State{$F \leftarrow \emptyset$ : {{\bf{thread private}} forbidden color set for $v$}}
\State{$W_{local} \leftarrow \emptyset$ : {{\bf{thread private}} vertices to be colored}}

\For{each $u \in$ $\vtxs$($v$)}\label{ln:kk:01}
\If{$c[u] \neq {\mathbf {-1}}$ {\bf and} $c[u] \notin F$}
\State{$F \leftarrow F \cup \{c[u]\}$}
\Else
\State{$W_{local} \leftarrow W_{local} \cup \{u\}$}\label{ln:kk:02}
\EndIf
\EndFor

\State{$col \leftarrow |\vtxs(v)| - 1$} \Comment{reverse first-fit coloring}\label{ln:rff1}
\For{each $u \in W_{local}$}
\While{$col \in F$}
\State{$col \leftarrow col - 1$}
\EndWhile
\State{$c[u] \leftarrow col$}\label{ln:rff2}
\State{$col \leftarrow col - 1$}\label{ln:rff3}
\EndFor
\EndFor
\end{algorithmic}
\label{alg:bgpc:n:color}
\end{algorithm}

\begin{table}
\scalebox{0.95}{
\begin{tabular}{l|r||r|r|r}
&&\multicolumn{3}{c}{Remaining $|W_{next}|$ after the first iteration}\\
Matrix-Graph&  $|V_{B}|$ &Alg.~\ref{alg:bgpc:n:color:ff} & Alg.~\ref{alg:bgpc:n:color:ff}  + reverse& Alg.~\ref{alg:bgpc:n:color} \\\hline
bone010 &  986,703 & 863,785 & 806,264 & 610,924 \\
coPapersDBLP& 540,486 & 409,621& 303,152 & 133,874
\end{tabular}}
\caption{{The number of uncolored (remaining) vertices after the first iteration for two graphs, obtained from matrices bone010 and coPapersDBLP, when Algorithms~\ref{alg:bgpc:n:color:ff} and~\ref{alg:bgpc:n:color} are used on 16 threads.}}
\label{tab:prelim}
\vspace*{-5ex}
\end{table}
\renewcommand{\baselinestretch}{1}

A drawback of the net-based conflict detection is the need of traversing all the nets for all iterations. For the vertex-based approach, it is sufficient to visit only the neighborhood of the vertices colored at the current iteration. However, without an intelligent net-marking technique in the coloring phase, it is not easy to restrict the neighborhood that needs to be traversed to identify all the conflicts. Hence, net-based conflict removal can be much faster than the vertex-based variant for the first few iterations. Although it can make the performance even worse for later iterations, in our experiments, $78\%$ of the runtime is observed to be used on the first iteration. That number goes up to $89\%$ for the first two iterations on average for eight graphs we used. Thus, attacking these first iterations would be enough.\looseness=-1

\begin{figure}
 \centering
 \includegraphics[width=.93\linewidth]{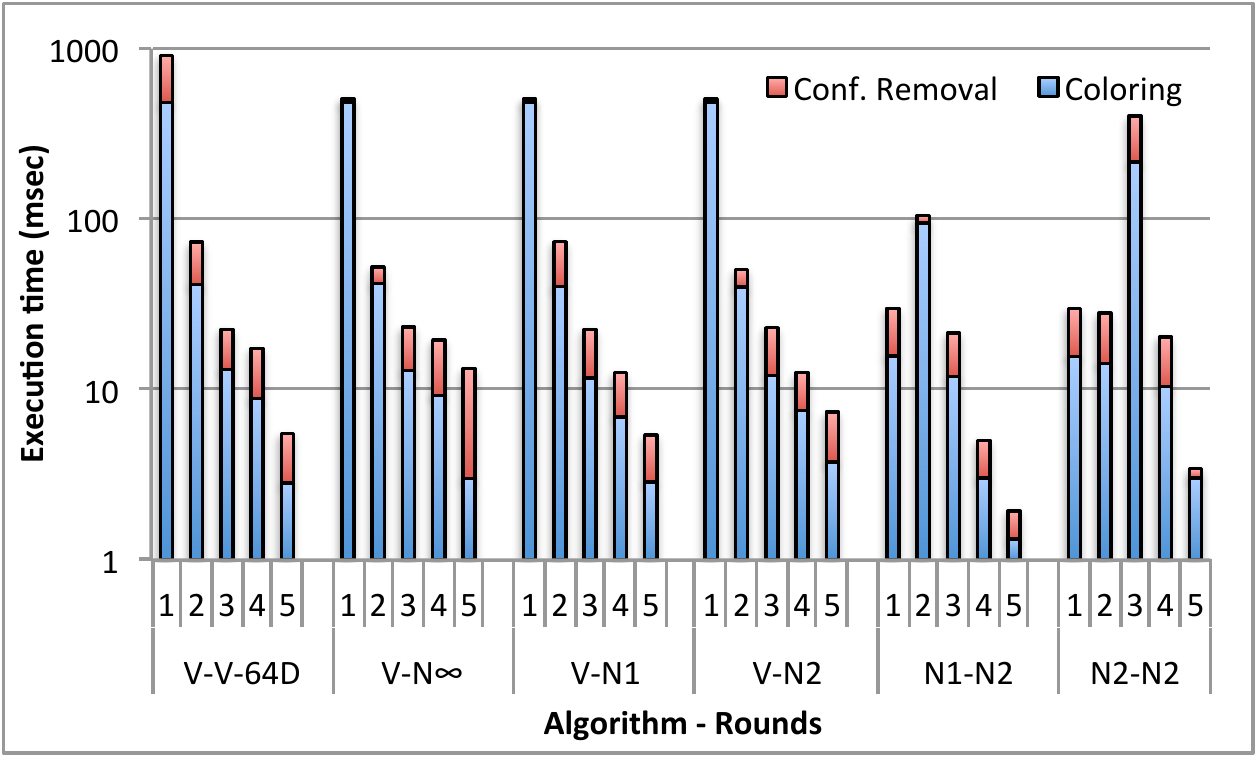}
\caption{{The execution times (in msec.) of each iteration for various algorithms while coloring coPapersDBLP with 16 threads.}}
\label{fig:round}
\end{figure}

Figure~\ref{fig:round} shows the execution times of each iteration of different algorithms while coloring coPapersDBLP with 16 threads. In the figure, an algorithm X-Y applies X-based coloring and Y-based conflict removal where the letter V and N denote vertex- and net-based, respectively. A number $n$ adjacent to the letter N denotes that the algorithm performs net-based approach for the first $n$ iterations and switch to the vertex-based approach. A more detailed explanation of the algorithms is given in Section~\ref{sec:exp}. The figure tells that: 1) most of the time is spent for the coloring, 2) most of the time is spent in the first iterations, 3) using net-based conflict removal at every iteration can make the performance worse~(V-N$\infty$), 4) using net-based coloring is a performance-wise good idea for the first iteration~(N1-N2), 5) using an additional net-based coloring at the second iteration is not useful~(N1-N2). The last observation can be obsolete if a better net-based~(or a hybrid) coloring approach is found.\looseness=-1
\vspace{1ex}
\noindent{\bf Implementation details:} For all the algorithms described above, the memories for the forbidden color set $F$ and the local vertex queues $W_{local}$ are allocated only once and simple arrays are used to realize them. Furthermore, these structures are never actually emptied or reset. For each thread, $F$ is repetitively used for different nets/vertices via different markers without any reset operation. Similarly, the local queue $W_{local}$ is emptied by only setting a local pointer to $0$.\looseness=-1

\section{Algorithms for Distance-2 Graph Coloring}\label{sec:D2GC}

The net-based approach can also be used for the D2GC problem. Due to the similarity between the problem definitions of BGPC and D2GC, the corresponding vertex- and net-based algorithms can be implemented along the lines of the bipartite graph partial coloring algorithms given above with a single difference: distance-1 neighbors must also be considered in the neighborhood as well. Here for completeness, we present the pseudo-codes for net-based D2GC coloring and conflict removal phases in Algorithms~\ref{alg:d2gc:n:color} and~\ref{alg:d2gc:n:conf}, respectively, but skip the vertex-based versions due to the space limitation. Since the input graph $G = (V, E)$ is unipartite, but not bipartite, instead of the $\nets(u)$ and $\vtxs(u)$, the notation $\nbor$($u$) will be used to denote the adjacency list of a vertex $u \in V$. However, for consistency, we will keep naming these greedier versions as net-based.\looseness=-1

\renewcommand{\baselinestretch}{0.9}
\begin{algorithm}
\caption{\textsc{D2GC-ColorWorkQueue-Net}}
\algorithmicrequire{ $G = (V, E)$: a graph, $c[.]$: an incomplete coloring.}\\
\algorithmicensure{ $c[.]$:  the (most) optimistic coloring array.}
\begin{algorithmic}[1]
\For{each $v \in V$ {\bf in parallel}} 
\State{$F \leftarrow \emptyset$ : {{\bf{thread private}} forbidden color set for $v$}}
\State{$W_{local} \leftarrow \emptyset$ : {{\bf{thread private}} vertices to be colored}}

\If{$c[v] \neq {\mathbf {-1}}$}\label{ln:d2:1}
\State{$F \leftarrow F \cup \{c[v]\}$}
\Else
\State{$W_{local} \leftarrow W_{local} \cup \{v\}$}
\EndIf\label{ln:d2:2}

\For{each $u \in$ $\nbor$($v$)}\label{ln:d2:kk:01}
\If{$c[u] \neq {\mathbf {-1}}$ {\bf and} $c[u] \notin F$}
\State{$F \leftarrow F \cup \{c[u]\}$}
\Else
\State{$W_{local} \leftarrow W_{local} \cup \{u\}$}
\EndIf
\EndFor

\State{$col \leftarrow |\nbor$($v$)$|$} \Comment{reverse first-fit coloring}\label{ln:d2:rff1}
\For{each $u \in W_{local}$}
\While{$col \in F$}
\State{$col \leftarrow col - 1$}
\EndWhile
\State{$c[u] \leftarrow col$}\label{ln:d2:rff2}
\State{$col \leftarrow col - 1$}\label{ln:d2:rff3}
\EndFor
\EndFor
\end{algorithmic}
\label{alg:d2gc:n:color}
\end{algorithm}

\begin{algorithm}
\caption{\textsc{D2GC-RemoveConflicts-Net}}
\algorithmicrequire{ $G = (V, E)$: a graph to color, $c[.]$: an optimistic coloring.}\\
\algorithmicensure{ $c[.]$: an incomplete coloring.}
\begin{algorithmic}[1]
\For{each $v \in V$ {\bf in parallel}} 
\State{$F \leftarrow \emptyset$ : {{\bf{thread private}} forbidden color set for $v$}}
\If{$c[v] \neq {\bf -1}$}\label{ln:d2:3}
\State{$F \leftarrow F \cup \{c[v]\}$}
\EndIf\label{ln:d2:4}
\For{each $u \in$ $\nbor$($v$)}
\If{$c[u] \neq {\bf -1}$}
\If{$c[u] \in F$}
\State{$c[u] \leftarrow \mathbf {-1}$}
\Else
\State{$F \leftarrow F \cup \{c[u]\}$}
\EndIf
\EndIf
\EndFor
\EndFor
\end{algorithmic}
\label{alg:d2gc:n:conf}
\end{algorithm}
\renewcommand{\baselinestretch}{1}

Unlike the BGPC algorithms, for D2GC, the threads visit actual vertices to be colored; this is why, both of the D2GC coloring and conflict removal algorithms first process the color of the visited vertices~(lines~\ref{ln:d2:1}--\ref{ln:d2:2} of Algorithm~\ref{alg:d2gc:n:color} and lines~\ref{ln:d2:3}--\ref{ln:d2:4} of Algorithm~\ref{alg:d2gc:n:conf}). This is necessary to handle the distance-1 neighbors which is the additional requirement for D2GC compared to BGPC. The same reverse first-fit policy is applied while coloring the vertices; the only difference is the candidate color is initialized with $|\nbor(v)|$ instead of $|\nbor(v)| - 1$~(as in D2GC) since the vertex assigned to a thread will also be colored by the thread requiring at least  $|\nbor(v)| + 1$ available colors~(including color 0).\looseness=-1

\section{Balancing color set cardinalities}\label{sec:bal}

As mentioned before, graph coloring has been frequently used to parallelize a large task with many sub-tasks. In our preliminary experiments, the (reverse) first-fit policy generated a few large color sets~(of small colors) and thousands of color sets with less than 2 elements for a real-life optimization problem. This result is in concordant with a comprehensive recent study focusing solely on balancing, parallel balancing heuristics, and their practical impacts on parallel computing~\cite{LuBalanced}. In fact, on a single multicore CPU, the performance reduction~(in FLOPS) may not hurt too much since most of the vertices, with small colors, can still be processed in parallel. However, the impact of the imbalance increases with the number of processors/cores. Furthermore, in most of the iterative algorithms, processing only a few vertices and updating the current solution can be harmful from the optimization perspective since this restricts the dimensions of the moves in the search space performed to reach a better solution.\looseness=-1
 
In this work, we experimented on cost-free and unsupervised balancing heuristics within the BGPC and D2GC algorithms proposed above. The straightforward choice would be keeping color set cardinalities dynamically throughout the execution; but this is expensive especially for large number of cores. Instead, the first proposed heuristic tries to keep the number of colors the same as much as possible
and the second one aggressively applies balancing hence increases the number of colors~(only around $10\%$ on average). The heuristics are given in Algorithms~\ref{alg:color:B1} and~\ref{alg:color:B2}  for the vertex-based approach. The net-based variants are also similar.\looseness=-1

\begin{algorithm}
\caption{\textsc{ColorWorkQueue-B1}}
\algorithmicrequire{ $G = (V, E)$, $W$: vertices to color, $\nbor$($.$): the neighborhood, $c[.]$: an incomplete coloring with no conflicts.}\\
\algorithmicensure{ $c[.]$:  an optimistic coloring.}
\begin{algorithmic}[1]
\State{$col_{max} \leftarrow 0$ : {\bf thread private}} \label{ln:b1}
\For{each $w \in W$ {\bf in parallel}} 
\State{\ldots} \Comment{lines 2-6 of Alg.~\ref{alg:color}}
\If{$w \bmod 2 = 0$}  
\State{$col \leftarrow col_{max}$} 
\While{$col \in F$}
\State{$col \leftarrow col - 1$}
\EndWhile
\If{$col = {\bf - 1}$}\label{ln:b1-s}
\State{$col \leftarrow col_{max} + 1$} 
\While{$col \in F$}
\State{$col \leftarrow col + 1$}
\EndWhile
\EndIf
\Else
\State{$col \leftarrow 0$} 
\While{$col \in F$}
\State{$col \leftarrow col + 1$}
\EndWhile
\EndIf
\State{$c[w] \leftarrow col$}
\State{$col_{max} = max(col_{max}, col)$}
\EndFor
\end{algorithmic}
\label{alg:color:B1}
\end{algorithm}

In the first balancing heuristic B1, each thread keeps track of the maximum color it uses ($col_{max}$ at line~\ref{ln:b1}). The threads employ the first-fit policy for the odd-numbered vertices (or nets) and otherwise, they employ the reverse first-fit policy starting from $col_{max}$. Unlike the original BGPC and D2GC algorithms, starting from $col_{max}$, instead of $|\nbor(w)| - 1$, necessitates a safety check~(line \ref{ln:b1-s}). If this is the case, the heuristic initiates a first-fit starting from $col_{max} + 1$. By performing alternating policies w.r.t. the vertex (or net) id, B1 hopes to distribute the colors evenly in the interval $[0, col_{max}]$. If there is no color between this interval, it extends the size of the interval.\looseness=-1

The second heuristic B2, given in Algorithm~\ref{alg:color:B2}, keeps a variable $col_{next}$ tIn addition to $col_{max}$ to start from the color search. The idea is the same: the heuristic wants to distribute the colors in between $[0, col_{max}]$ but increments the color to start by one for each vertex/net. To aggressively favor large color numbers and focus the later colors in the interval more, the minimum color to start is set to  $col_{max} / 3  + 1$~(the last line of Alg.~\ref{alg:color:B2}).  However, filling these color sets with more vertices increases the probability of them being in a forbidden-color array. Thus, more colors are expected to appear during the course of execution due to the conflicting nature of balancing and using less number of colors.\looseness=-1

\renewcommand{\baselinestretch}{0.9}
\begin{algorithm}
\caption{\textsc{ColorWorkQueue-B2}}
\algorithmicrequire{ $G = (V, E)$, $W$: vertices to color, $\nbor$($.$): the neighborhood, $c[.]$: an incomplete coloring with no conflicts.}\\
\algorithmicensure{ $c[.]$:  an optimistic coloring.}
\begin{algorithmic}[1]
\State{$col_{max} \leftarrow 0$ : {\bf thread private}} 
\State{$col_{next} \leftarrow 0$ : {\bf thread private}} 
\For{each $w \in W$ {\bf in parallel}} 
\State{\ldots} \Comment{lines 2-6 of Alg.~\ref{alg:color}}
\State{$col \leftarrow col_{next}$} 
\While{$col \in F$}
\State{$col \leftarrow col  + 1$}
\EndWhile
\If{$col > col_{max}$}
\State{$col \leftarrow 0$} 
\While{$col \in F$}
\State{$col \leftarrow col + 1$}
\EndWhile
\EndIf
\State{$c[w] \leftarrow col$}
\State{$col_{max} = max(col_{max}, col)$}
\State{$col_{next} = min(col + 1, col_{max} / 3  + 1)$}
\EndFor
\end{algorithmic}
\label{alg:color:B2}
\end{algorithm}
\renewcommand{\baselinestretch}{1}

\section{Experiments}\label{sec:exp}
All the experiments in the paper are performed on a single machine running on 64 bit CentOS 6.5 equipped with 64GB RAM and a dual-socket Intel Xeon E7-4870 v2 clocked at 2.30 GHz where each socket  has 15 cores~(30 in total). For the multicore implementations, we used OpenMP and all the codes are compiled with {\tt gcc 4.9.2} with the {\tt -O3} optimization flag enabled. For each problem, we experimented on eight different algorithms which are combinations of the heuristics given in Sections~\ref{sec:BGPC} and \ref{sec:D2GC}. For fairness, all the algorithms are implemented within the \CP environment using the same data structures as much as possible. All the algorithms are summarized below:

\begin{itemize}
\footnotesize
\item {{\bf V-V:} vertex-based coloring and conflict removal with first-fit policy. This is the default implementation of \CP for BGPC. For D2GC, \CP does not have a parallel implementation but a sequential one exists. We implemented the parallel version based on the BGPC algorithm by adding the corresponding statements for distance-1 neighbors.}

\item {{\bf V-V-64:} Same as V-V but the chunk-size for dynamic scheduling of OpenMP threads are set to 64.}

\item {{\bf V-V-64D:} In \CP's conflict removal, a conflicting vertex is immediately added to the shared work queue of the next iteration. Unlike V-V64, this algorithm performs a lazy construction by using private queues for each thread that are combined at the end of each iteration.}

\item {{\bf V-N$\infty$}, {\bf V-N1} and {\bf V-N2}: Vertex-based coloring~(64D) with net-based conflict removal in all, the first, and the first two iterations, respectively. After that, the algorithms switch to vertex-based~(64D)  conflict removal.}

\item {{\bf N1-N2} and {\bf N2-N2}: Similarly, these two algorithms use net-based coloring in the first and first-two iterations, respectively, and both use net-based conflict removal only in the first two iterations. The algorithms switch to the vertex-based~(64D) variants after that.}

\end{itemize}

\renewcommand{\baselinestretch}{0.95}
\begin{table*}[htbp]
\begin{center}
\scalebox{0.93}{
\begin{tabular}{l||rrr|rr||rr|rr||r}          
&\multicolumn{5}{c||}{Properties}&\multicolumn{4}{c||}{Sequential BGPC V-V} & \\\cline{1-10}
&&&&\multicolumn{2}{c||}{Column deg.} &\multicolumn{2}{c|}{Natural}&\multicolumn{2}{c||}{Smallest Last} & \\
Matrix&\#rows&\#cols&\#nnz&max.& Std. dev.&Exec. time&\#colors&Exec. time&\#colors&Used\\\hline
20M\_movielens & 26,744 & 138,493 &  20,000,263 & 67,310 & 3,085.81 & 587.15 & 70,815&1,236.33 & 68,077& \checkmark
 $\times$\\	
af\_shell~\cite{patwary} & 1,508,065 & 1,508,065 & 27,090,195 & 35 & 1.00 & 3.39 &50 &4.13 & 45& \checkmark
 \checkmark\\	
bone010~\cite{patwary}    & 986,703  & 986,703  & 36,326,514 & 63 & 7.61 & 4.28& 132 & 6.86 & 110& \checkmark
 \checkmark\\	
channel~\cite{LuBalanced}    & 4,802,000 & 4,802,000 & 42,681,372 & 18 & 1.00 & 2.57  & 39 & 4.75 & 36& \checkmark
 \checkmark\\	
coPapersDBLP~\cite{LuBalanced} &540,486 & 540,486  & 15,245,729 & 3,299 & 66.23 & 6.73 & 3,321 &9.68 & 3,300& \checkmark
 \checkmark\\	
HV15R~\cite{bergen}     & 2,017,169 & 2,017,169 & 283,073,458 & 484 & 53.95 & 66.94 & 508 &87.01 & 484& \checkmark
$\times$\\	
nlpkkt120~\cite{patwary}  & 3,542,400 & 3,542,400 & 50,194,096  & 28 & 3.00 & 4.22 & 59 &7.88 & 49 &  \checkmark
 \checkmark\\	
uk-2002~\cite{LuBalanced}  & 18,520,486 & 18,520,486 & 298,113,762 & 2,450 & 27.51& 32.66 & 2,450 & 41.23 & 2,450 &  \checkmark
$\times$\\	
\end{tabular}
}
\caption{\scriptsize{Graphs/matrices used in the experiments: columns 2-4 are the numbers of rows, columns, and nonzeros, respectively. The next two columns are the maximum number of nonzeros in a column and the standard deviation of the nonzero distribution. Columns 7-8 show the execution time of the  sequential BGPC algorithm and the average number of colors when the natural row order is employed. The next columns do the same for the smallest-last order implemented in \CP to reduce the number of distinct colors. The last column show if the matrix is used in BGPC and D2GC experiments, respectively. The ordering time is not included in the table. Moreover, since the executions are sequential, a conflict detection phase is not performed.}}\label{tbl:graphs}
\end{center}
\end{table*}
\renewcommand{\baselinestretch}{1}

The experiments are performed on eight graphs  given in Table~\ref{tbl:graphs} which are generated from their corresponding UFL matrices~\cite{UFL}. Seven out of the eight graphs have been taken from the coloring and related parallel computing literature~\cite{LuBalanced,patwary,bergen}. We also included a matrix from MovieLens dataset~\cite{movielens}, 20M\_movielens, since matrix decomposition, and our preliminary experiments on these matrices, is the application that motivated us for this study. For BGPC, we colored the columns of these matrices where the rows are considered as the nets. For D2GC, we used 5 of 8 structurally symmetric matrices. This is denoted in the last column of the table.\looseness=-1

\subsection{Experiments for bipartite graph partial coloring}
\begin{figure*}
  \centering
\subfigure[20M\_movielens]{\includegraphics[width=.49\linewidth]{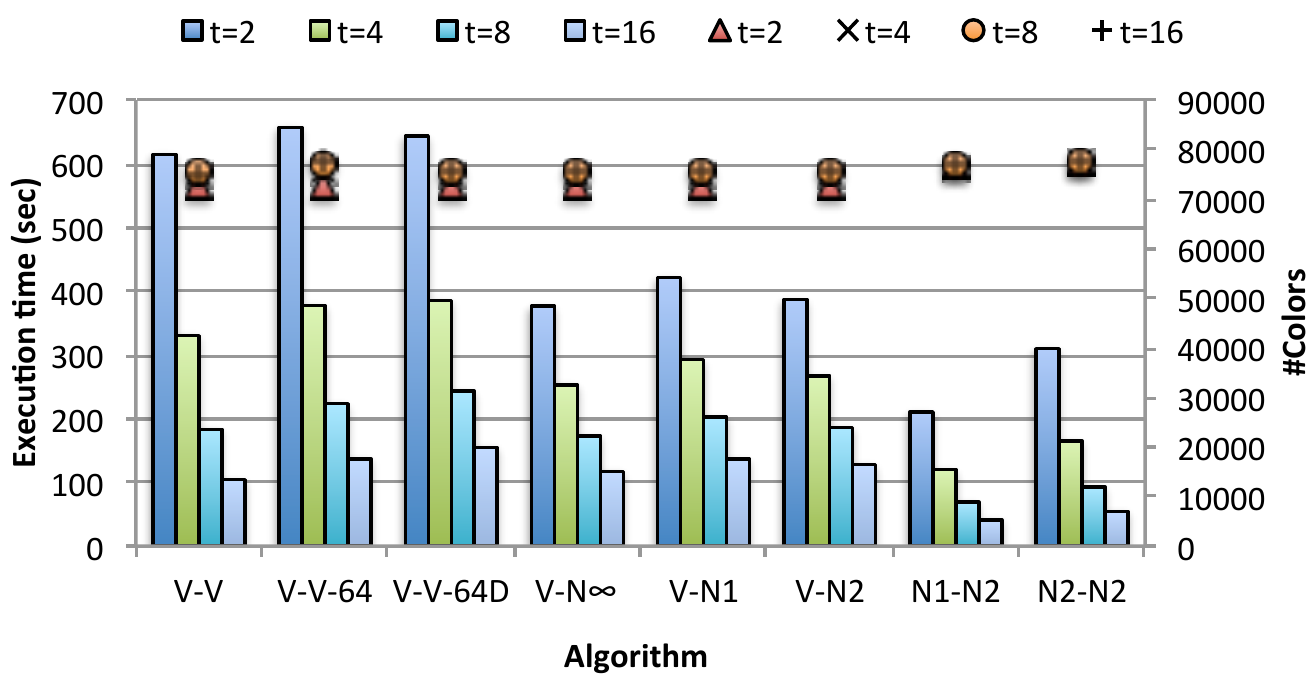}}
\subfigure[af\_shell]{\includegraphics[width=.49\linewidth]{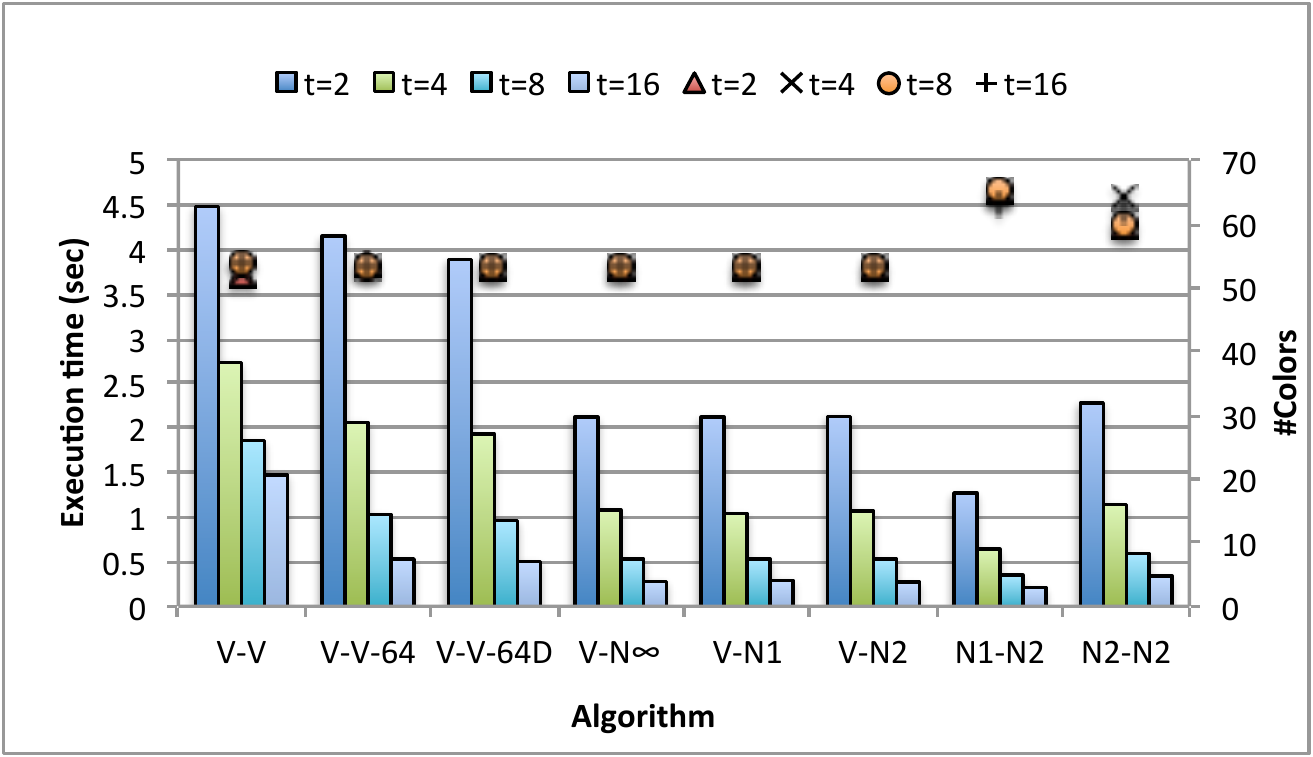}}
\subfigure[bone010]{\includegraphics[width=.49\linewidth]{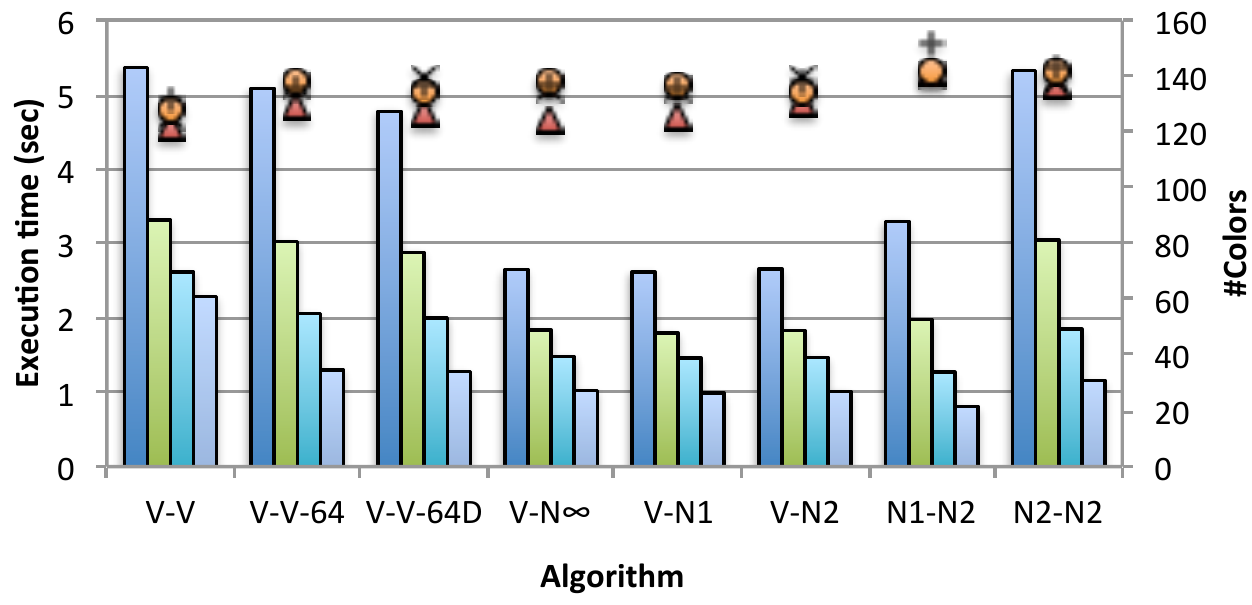}}
\subfigure[channel]{\includegraphics[width=.49\linewidth]{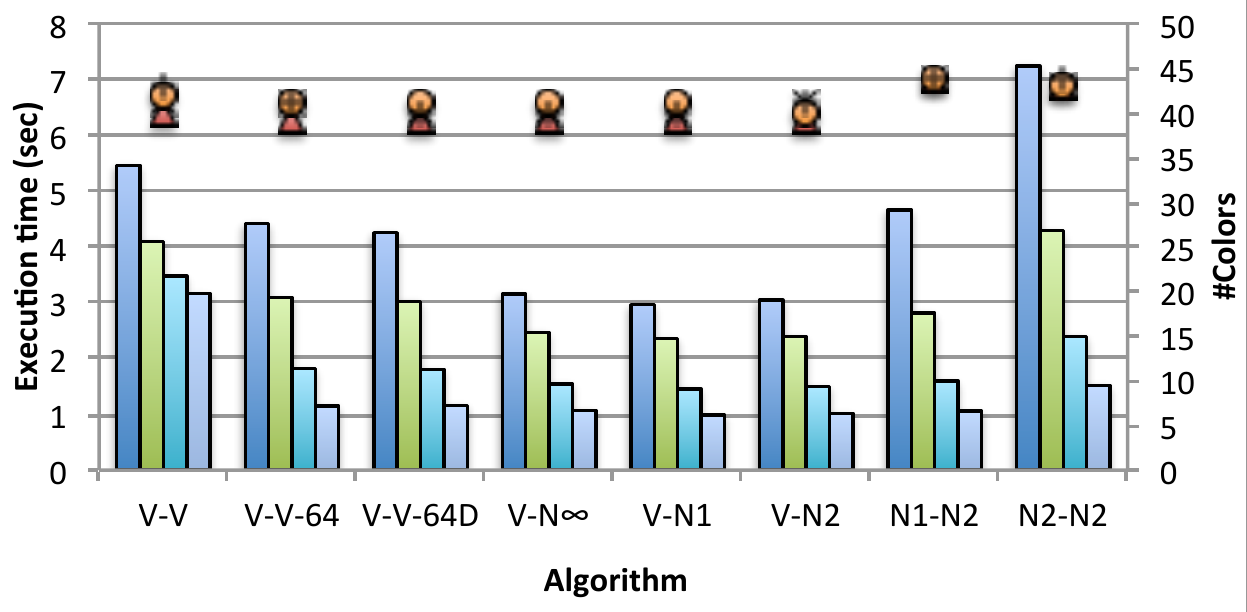}}
\subfigure[coPapersDBLP]{\includegraphics[width=.49\linewidth]{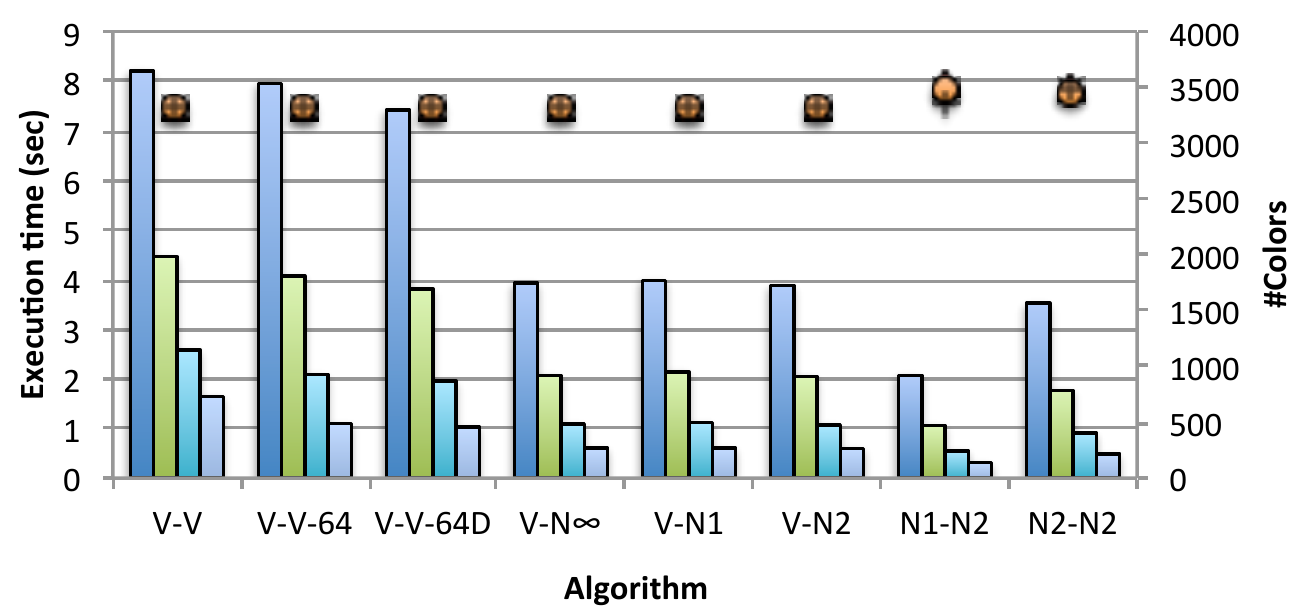}}
\subfigure[HV15R]{\includegraphics[width=.49\linewidth]{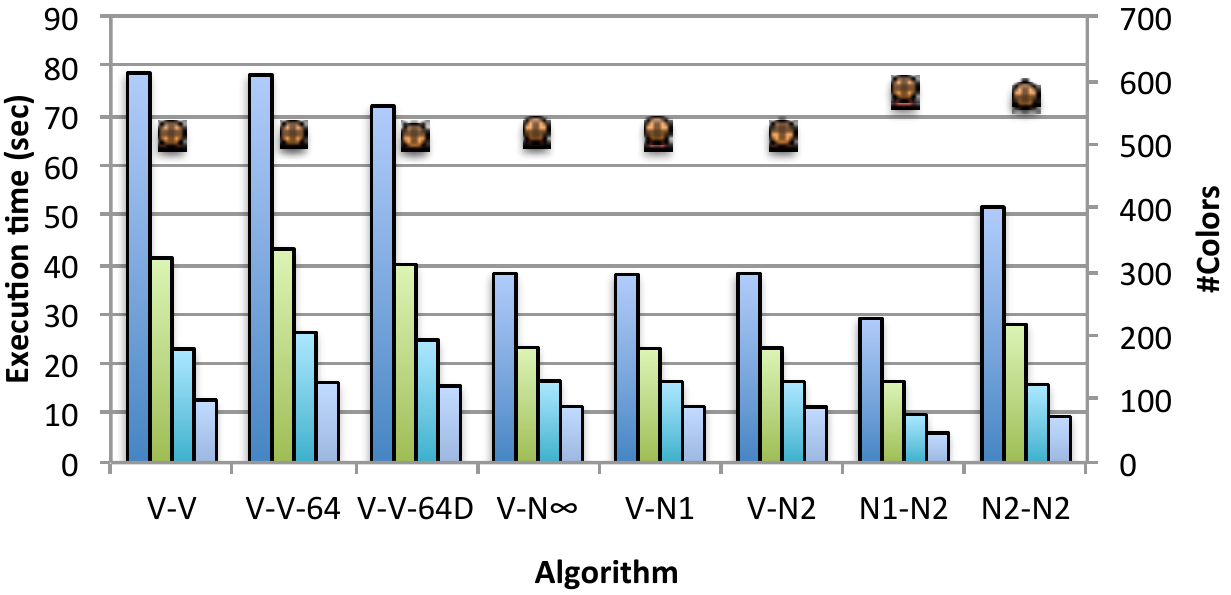}}
\subfigure[nlpkkt120]{\includegraphics[width=.49\linewidth]{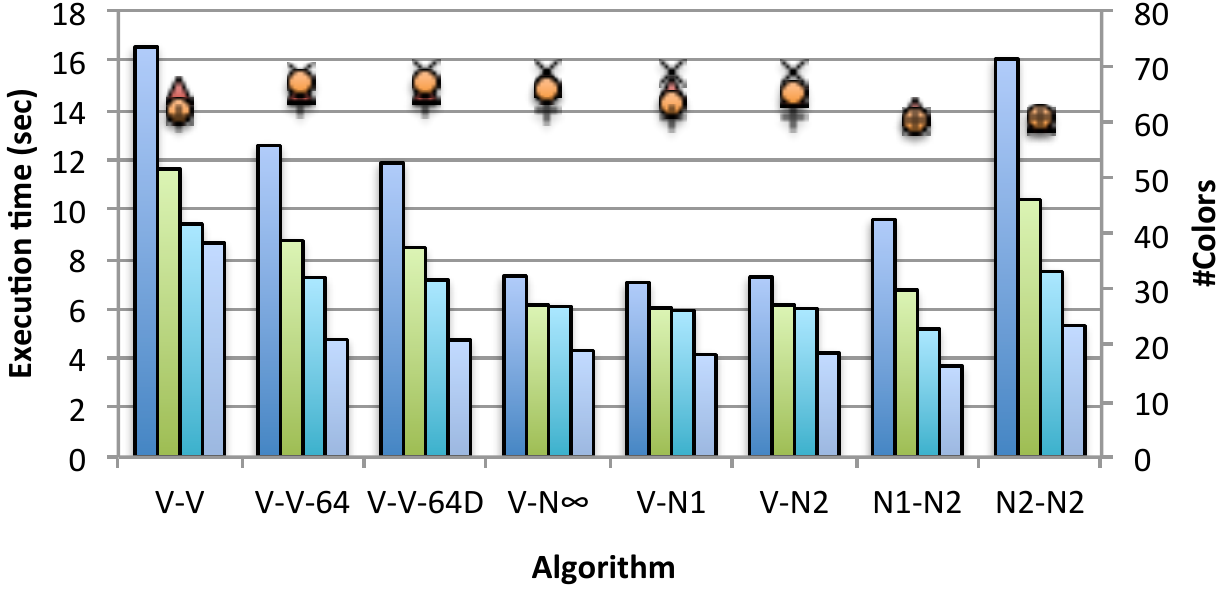}}
\subfigure[uk\_2002]{\includegraphics[width=.49\linewidth]{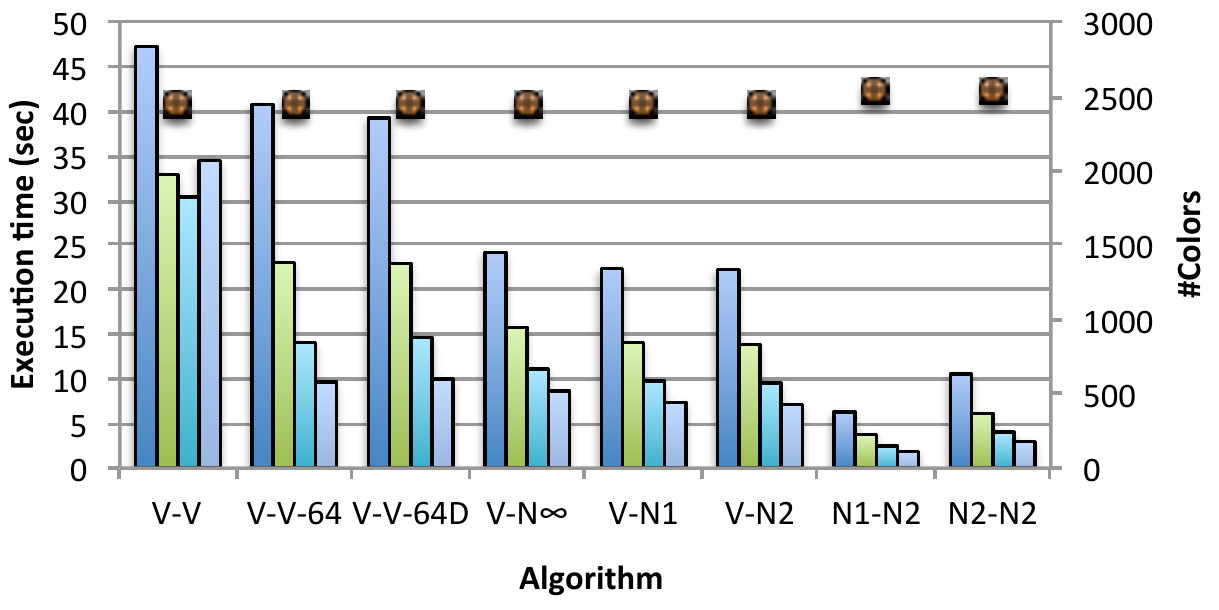}}
\caption{The execution times~(left axis) on 2, 4, 8, 16 threads, respectively, and the number of colors~(right axis) for all the matrices and algorithms.}
\label{fig:BGPC_tables}
\end{figure*}

The execution times of BGPC algorithms for each matrix as well as the number of distinct colors are given in Figure~\ref{fig:BGPC_tables} and the results are summarized in Table~\ref{tbl:smallestbipartitetable}. When the natural vertex order is used, compared to sequential \CP implementation of BGPC, i.e., V-V, one can obtain $6.01\times$ speed-up on 16 threads with $1\%$ increase on the number of colors~(V-N2). When the net-based coloring is employed for one iteration~(N1-N2), the speedup increases to $11.38\times$ with a small, $8\%$ increase on the number of colors. These algorithms are $2.17\times$ and $4.12\times$, respectively, faster than the parallel BGPC in \CP on 16 threads.\looseness=-1

We also used compared the results when the smallest-last order in \CP is employed. As Table~\ref{tbl:graphs} shows, this ordering indeed reduces the number of colors for most of the cases. The results of these experiments are summarized in Table~\ref{tbl:smallestbipartitetable}. Since the sequential \CP execution for this ordering is slower than that of the natural ordering, the speedups increase: compared to sequential V-V, the algorithms V-N2 and N1-N2 are $10.09\times$ and  $16.76\times$ faster, respectively, with 16 threads. Compared to parallel V-V, on 16 threads, N1-N2 is $4.43\times$ faster with $9\%$ increase on the number of colors used.\looseness=-1

\begin{table}
\scalebox{0.9}{
\begin{tabular}{l||r||r|r|r|r||r}
&Avg. \#colors&\multicolumn{4}{c||}{}&Speedup\\
&normalized&\multicolumn{4}{c||}{Speedup over sequential V-V}& over V-V\\
Algorithm&w.r.t. V-V&	$t=2$	&$t=4$&	$t=8$&	$t=16$& for $t=16$	\\\hline
V-V&			1.00& 	0.74&	1.24&	1.88&	2.76& 	1,00\\\hline	
V-V-64&		1.01&	0.81&	1.40&	2.36&	4.00&	1,45\\
V-V-64D&	1.01&	0.85&	1.46&	2.41&	4.05&	1,47\\\hline
V-N$\infty$&	1.01&	1.47&	2.34&	3.65&	5.84&	2,11\\
V-N1& 		1.01&	1.48&	2.35&	3.64&	5.85&	2,11\\
V-N2& 		1.01&	1.49&	2.37&	3.71&	6.01&	{\bf 2,17}\\\hline
N1-N2& 		1.08&	2.39&	4.24&	7.17&	11.38& 	{\bf 4,12}\\
N2-N2& 		1.07&	1.44&	2.63&	4.57&	7.50&	2,71
\end{tabular}
}
\caption{\scriptsize{The average speedups over sequential and parallel V-V on 16 threads and the increase on the number of colors when the natural ordering of the columns is used. The numbers are the geometric means of the individual results for each matrix.}}
\label{tbl:naturalbipartitetable}
\end{table}

\begin{table}
\scalebox{0.9}{
\begin{tabular}{l||r||r|r|r|r||r}
&Avg. \#colors&\multicolumn{4}{c||}{}&Speedup\\
&normalized&\multicolumn{4}{c||}{Speedup over sequential V-V}& over V-V\\
Algorithm&w.r.t. V-V&	$t=2$	&$t=4$&	$t=8$&	$t=16$& for $t=16$	\\\hline
V-V&1.00&	0.93&	1.65&	2.81&	3.78& 	1,00\\\hline	
V-V-64&1.01&	0.99&	1.89&	3.55&	6.41& 	1.70\\
V-V-64D&0.99&	1.04&	1.99&	3.75&	6.86&	1.81\\\hline
V-N$\infty$&1.00&	1.62&	3.01&	5.41&	9.20&	2.43\\	
V-N1& 1.01&	1.71&	3.19&	5.83&	10.07&	2.66\\
V-N2& 0.99&	1.72&	3.21&	5.87&	10.09&	{\bf 2.67}\\\hline	
N1-N2& 1.09&	3.47&	6.26&	10.82&	16.76&	{\bf 4.43}\\
N2-N2& 1.10&	2.24&	4.04&	6.94&	11.19&	2.96\\
\end{tabular}
}
\caption{\scriptsize{The average speedups over sequential and parallel V-V on 16 threads and the increase on the number of colors when smallest-last order of the columns is used. The numbers are the geometric means of the individual results for each matrix.}}			
\label{tbl:smallestbipartitetable}
\end{table}

\subsection{Experiments for distance-2 graph coloring}

For D2GC, we have experimented on the five of eight matrices in our data set as explained above. We've selected four algorithms which obtained promising results in the BGPC experiments. The results are presented in~Table~\ref{tbl:d2results}. Similar to BGPC, 16-thread V-N1 and N1-N2 is $8.97\times$ and $13.2\times$ faster than sequential V-V with only $4\%$ and $9\%$ increase in color counts, respectively. V-V-64D is used to normalize the 16-thread speedups since all the algorithms employ the 64D option. When the improvement of chunk size and lazy work-queue construction is removed, the optimism in N1-N2 obtains $2\times$ performance on 16-threads with only around $5\%$ increase on the number of distinct colors.\looseness=-1

\begin{table}
\begin{center}
\scalebox{0.9}{
\begin{tabular}{l||c||rrrr||r} 
& Color &\multicolumn{4}{c||}{}&	Speedup over\\
& w.r.t. &\multicolumn{4}{c||}{Speedup over sequential V-V}&	 V-V-64D\\
Algorithm & V-V &	$t=2$&	$t=4$&	$t=8$&	$t=16$&for $t=16$\\\hline
V-V-64D&	1.04&	1.38&	2.18&	3.46&	6.11&	1.00\\\hline
V-N1&	1.04&	2.32&	3.38&	5.22&	8.97&	1.39\\
V-N2	&1.04&	2.27&	3.37&	5.24&	8.87&	1.37\\\hline
N1-N2&	1.09&	2.49&	4.44&	7.85&	13.20&	2.00
\end{tabular}
}
\caption{\scriptsize{The average speedups over sequential V-V and parallel V-V-64D on 16 threads and the increase on the number of colors~(over V-V) when the natural ordering of the columns is used. The numbers are the geometric means of the individual results for each matrix. The results are the averages of 10 experiments for each matrix-algorithm-thread triplet.}}
\label{tbl:d2results}
\end{center}
\vspace*{-4ex}				
\end{table}

\subsection{Experiments on balancing}
\begin{figure*}
\begin{center}
\subfigure[Balancing coPapersDBLP for V-N2]{\includegraphics[width=.45\linewidth]{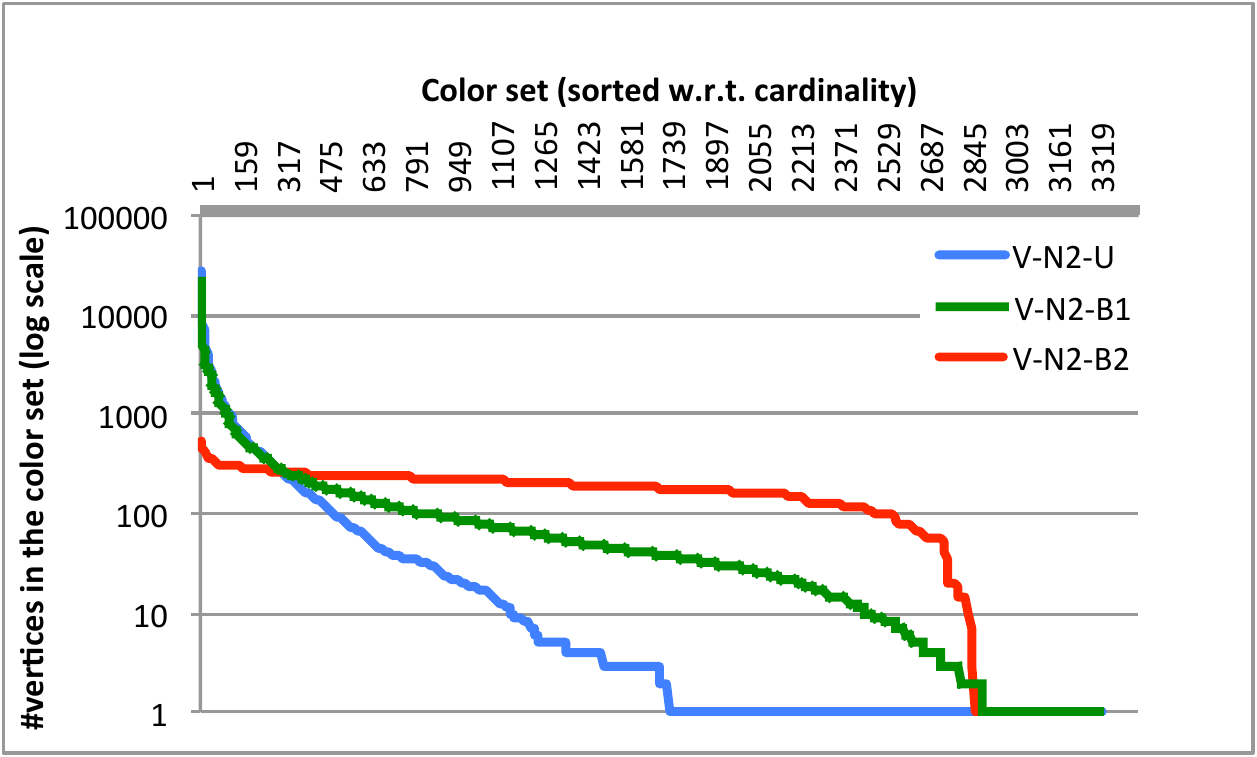}}
\subfigure[Balancing coPapersDBLP for N1-N2]{\includegraphics[width=.45\linewidth]{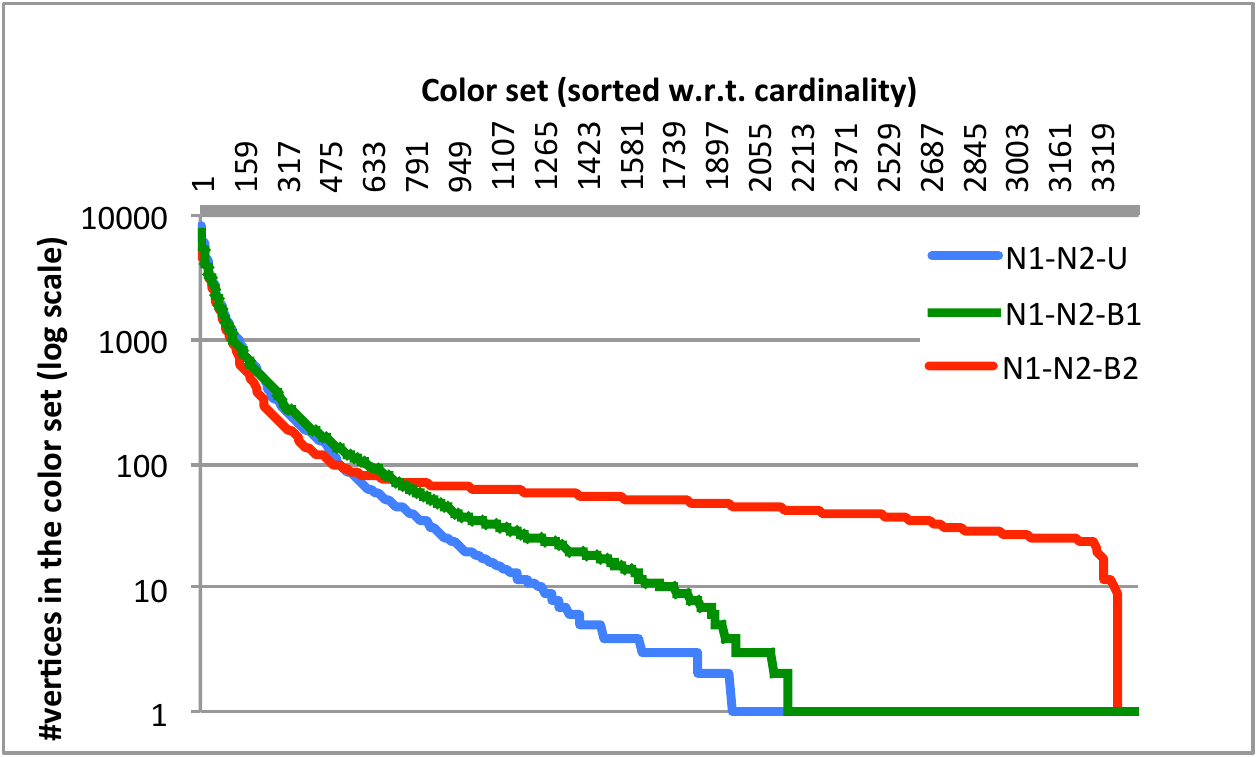}}
\caption{\scriptsize{Impact of balancing heuristics, B1 and B2, on the color set cardinalities and the number of color sets for BGPC algorithms parallel V-N2~(left) and N1-N2~(right) on 16-threads for coPapersDBLP.}}
\label{fig:balancefig}
\end{center}
\vspace*{-4ex}
\end{figure*}

The impact on balancing heuristics B1 and B2 are presented in~Table~\ref{tbl:balanced} for BGPC experiments.  
The heuristics are applied to V-N2 and N1-N2 and the results are compared with their original implementation. Experimental results show that, applying these heuristics is for free, i.e., there is no computational overhead as expected. For B1, the standard deviation of the color cardinalities decreases $0.69\times$ and $0.84\times$ when applied to V-N2 and N1-N2, respectively, on the expense of  $4\%$ color increase. For B2, which aggressively tries to reduce the number of colors, the standard deviation decreases $0.25\times$ and $0.62\times$ with around $9\%$ and $13\%$ increase on the number of colors for V-N2 and N1-N2, respectively. To better visualize the impact of these balancing heuristics, Figure~\ref{fig:balancefig} shows the distribution of color set cardinalities for the original and balanced executions of  V-N2 and N1-N2 on coPapersDBLP.\looseness=-1

\begin{table}

\begin{center}
\scalebox{0.9}{
\begin{tabular}{l||rrrr} 

&\multicolumn{4}{c}{Normalized w.r.t. X-N2}\\\hline
& Coloring &	$\#$Color  & Average  & Std.\\
Algorithm&  time & sets & card. & Dev. \\\hline
V-N2-U & 1.00  & 	1.00& 	1.00	& 1.00\\
V-N2-B1 &	0.95& 	1.04& 	0.96& 	0.69\\
V-N2-B2& 0.95& 	1.13& 	0.89& 	0.25\\\hline
N1-N2-U & 1.00& 	1.00& 	1.00& 	1.00 \\
N1-N2-B1 & 0.99& 	1.04& 	0.96& 	0.84 \\
N1-N2-B2 & 0.99& 	1.09& 	0.91& 	0.62 \\
\end{tabular}
}
\caption{\scriptsize{Impact of balancing heuristics, B1 and B2, on the color set cardinalities and the number of color sets for parallel BGPC algorithms V-N2 and N1-N2 on 16 threads. Results are normalized with the original unbalanced algorithms denoted with -U.}}
\label{tbl:balanced}
\vspace*{-6ex}
\end{center}				

\end{table}

\section{Related work}\label{sec:related}

Coloring has mostly been investigated for distance-1 coloring,
but most ideas can be ported to other variants. Since graph coloring
is NP-Complete~\cite{matula_SL} and hard to
approximate~\cite{Zuckerman} in most of its variants,
the vertices are
greedily colored one after another, and the lowest available color for
a vertex is selected. Such an algorithm produces a coloring with
less than $1+\Delta$ in distance-1 coloring. Though to avoid the
worst case, it is common to carefully choose the order in which
the vertices are processed~\cite{GMP05} using either a static ordering
~\cite{Matula1983,Welsh01011967}, or dynamic
ordering~\cite{Brelaz1979}.\looseness=-1

Earlier coloring algorithms~\cite{ABC94,JP93,GJP96},
are based generating maximum independent sets in parallel via algorithms such as~\cite{Luby86}. Though, recent techniques optimistically color the vertices in parallel 
assuming that a valid coloring will be generated and then verify
the validity of the coloring. One of the neighbor vertices
that are of the same color is tagged to be colored in the next
iteration of the algorithm. This technique was successfully applied
on distributed memory
machine~\cite{Boman05-EuroPar,BGMBC-jpdc,SSC11-HiPC,SSC14-arxiv},
including for BGPC and D2GC~\cite{Bozdag05-HPCC,Bozdag10-SISC}. The algorithm was investigated also on shared
memory, multicore and manycore architectures~\cite{Catalyurek12-ParCo,Gebremedhin_parallelgraph,patwary,Gebremedhin02paralleldistance-k,Deveci2016} and on hybrid
MPI~+~OpenMP systems~\cite{SSC12-PCO}. 
One common point of~\cite{Bozdag05-HPCC,Bozdag10-SISC} and the proposed work is that
the conflict removal phase of D2GC has been performed around middle vertices which is similar to
the net-based conflict removal. Nevertheless, the authors studied D2GC in the distributed setting and 
applied the approach for all iterations.\looseness=-1


The balanced graph coloring problem has been studied in the literature from different aspects;
from theoretical perspective, the term ``equitable" is used for the colorings where the color 
set cardinalities differ at most one~\cite{Meyer,Hajnal}. The most comprehensive study
from the parallel computing perspective is recently introduced by Lu~et~al.~\cite{LuBalanced}.
In this work, we follow a similar approach but mostly aim to devise costless and online balancing heuristics that can 
be applied to the parallel greedy graph coloring algorithms.\looseness=-1

\section{Conclusion and future work}\label{sec:conc}

In this work, we proposed novel, greedier and more optimistic parallel algorithms for parallel BGPC and D2GC. We also proposed two costless balancing heuristics that can be applied to both BGPC and D2GC, as well as other coloring variants, to balance the color set cardinalities and improve the impact of the coloring on the real application to be parallelized. The results show that the proposed techniques are useful in practice and improves the performance and the {\em goodness} of the coloring.\looseness=-1

The proposed techniques are suitable for GPUs and Intel Xeon Phi architectures which can be considered as a future work. In fact, the task sizes in the vertex-based approach, i.e., the neighborhood sizes, deviates much more compared to that of the net-based approach, i.e., number of vertices adjacent to a net,  which can be a comfort while parallelizing the coloring algorithms on manycore architectures. We also believe that a better net-based coloring and a better cost-free, self-balancing heuristic worth to investigate since their impact will be significant as the experimental results imply. 
  
\section*{Acknowledgments}
Kamer Kaya was supported by T{\"{U}}B\.{I}TAK BIDEB 2232 program under grant number 115C018.

\begin{small}
\renewcommand{\baselinestretch}{0.9}


\renewcommand{\baselinestretch}{1}
\end{small}

\end{document}